\documentstyle[12pt,aaspp4]{article}

\lefthead{Rey et al.}
\righthead{GALEX M31 Globular Clusters}

\received{2004 April 15}
\begin{document}
\title{{\sl GALEX} Ultraviolet Photometry of Globular Clusters in M31}

\author{Soo-Chang Rey\altaffilmark{1, 2, 3, 13}, R. Michael Rich\altaffilmark{1, 4}, 
Young-Wook Lee\altaffilmark{2}, Suk-Jin Yoon\altaffilmark{5}, Sukyoung K. Yi\altaffilmark{5}, 
Luciana Bianchi\altaffilmark{6}, Young-Jong Sohn\altaffilmark{2}, Peter G. Friedman\altaffilmark{3}, 
Tom A. Barlow\altaffilmark{3}, Yong-Ik Byun\altaffilmark{2}, Jose Donas\altaffilmark{7}, 
Karl Forster\altaffilmark{3}, Timothy M. Heckman\altaffilmark{8}, Myungkook J. Jee\altaffilmark{8}, 
Patrick N. Jelinsky\altaffilmark{9}, Suk-Whan Kim\altaffilmark{2}, Jae-Woo Lee\altaffilmark{10}, 
Barry F. Madore\altaffilmark{11}, Roger F. Malina\altaffilmark{7}, D. Christopher Martin\altaffilmark{3}, 
Bruno Milliard\altaffilmark{7}, Patrick Morrissey\altaffilmark{3}, Susan G. Neff\altaffilmark{12},
Jaehyon Rhee\altaffilmark{2, 3}, David Schiminovich\altaffilmark{3}, Oswald H. W. Siegmund\altaffilmark{9}, 
Todd Small\altaffilmark{3}, Alex S. Szalay\altaffilmark{8}, Barry Y. Welsh\altaffilmark{9}, 
and Ted K. Wyder\altaffilmark{3}}

\altaffiltext{1}{equal first authors  (screy@srl.caltech.edu, rmr@astro.ucla.edu)}
\altaffiltext{2}{Center for Space Astrophysics,
Yonsei University, Seoul 120-749, Korea}

\altaffiltext{3}{California Institute of Technology, MC 405-47, 1200 East
California Boulevard, Pasadena, CA 91125}

\altaffiltext{4}{Department of Physics and Astronomy, University of
California, Los Angeles, CA 90095}

\altaffiltext{5}{Department of Physics, University of Oxford,
Oxford OX1 3RH, UK} 

\altaffiltext{6}{Center for Astrophysical Sciences, The Johns Hopkins
University, 3400 N. Charles St., Baltimore, MD 21218}

\altaffiltext{7}{Laboratoire d'Astrophysique de Marseille, BP 8, Traverse
du Siphon, 13376 Marseille Cedex 12, France}

\altaffiltext{8}{Department of Physics and Astronomy, The Johns Hopkins
University, Homewood Campus, Baltimore, MD 21218}

\altaffiltext{9}{Space Sciences Laboratory, University of California,
Berkeley, 601 Campbell Hall, Berkeley, CA 94720}

\altaffiltext{10}{Department of Astronomy and Space Sciences,
Sejong University, Seoul 143-747, Korea}

\altaffiltext{11}{Observatories of the Carnegie Institution of Washington,
813 Santa Barbara St., Pasadena, CA 91101}

\altaffiltext{12}{Laboratory for Astronomy and Solar Physics, NASA Goddard
Space Flight Center, Greenbelt, MD 20771}

\altaffiltext{13}{Department of Astronomy and Space Sciences, Chungnam National University,
Daejeon 305-764, Korea}

\begin{abstract}

We present ultraviolet photometry for globular clusters (GCs) in M31 from 15 deg$^{2}$ of imaging
using the {\sl Galaxy Evolution Explorer} (GALEX). We detect 200 and 94 GCs with certainty in the 
near-ultraviolet (NUV; 1750 $-$ 2750 \AA) and far-ultraviolet (FUV; 1350 $-$ 1750 \AA) bandpasses, 
respectively. Our rate of detection is about 50\% in the NUV and 23\% in the FUV, to an approximate 
limiting $V$ magnitude of 19. Out of six clusters with [Fe/H]$>-1$ seen in the NUV, none is detected in 
the FUV bandpass. Furthermore, we find no candidate metal-rich clusters with significant FUV flux, 
because of the contribution of blue horizontal-branch (HB) stars, such as NGC 6388 and NGC 6441, 
which are metal-rich Galactic GCs with hot HB stars. We show that our GALEX photometry follows 
the general color trends established in previous UV studies of GCs in M31 and the Galaxy. 
Comparing our data with Galactic GCs in the UV and with population synthesis models, we suggest 
that the age range of M31 and Galactic halo GCs are similar.   
\end{abstract}


\section{Introduction}
The globular cluster (GC) system of M31 has been the subject of intensive study, 
due to the similarity of M31 to the Milky Way, its proximity, and the large numbers of 
cataloged clusters. Because the M31 GCs can be resolved into stars with $Hubble Space Telescope$ (HST) imaging 
(e.g., Ajhar et al. 1996; Rich et al. 1996; Fusi Pecci et al. 1996), it is possible to explore how
integrated colors depend on the stellar content of the GCs and to do so in a galaxy that
overall resembles the Milky Way.  The general properties of the M31 GC system are
described in recent reviews by Rich (2003), Jablonka (2002), and Barmby (2002).  
A recent optical/infrared catalog is published in Galleti et al. (2004).

Previous ultraviolet (UV) missions obtained photometry for Galactic GCs, beginning with the 
{\sl Orbiting Astronomical Observatory} (OAO2), the {\sl Astronomical Netherlands Satellite} (ANS),
the {\sl International Ultraviolet Explorer} (IUE), and the {\sl Ultraviolet Imaging Telescope} (UIT)
(Dorman, O'Connell, \& Rood 1995 and references therein). The first UV imagery of the 
M31 GC system was obtained using the UIT (Bohlin et al. 1993). However, the UIT images from 
photographic film detected 43 GCs in the near-ultraviolet (NUV), while only four clusters had solid 
far ultraviolet (FUV) detections.  These early observations allowed Bohlin et al. to compare 
the UV properties of a small sample of M31 GCs with those of Galactic halo GCs and to conclude 
that the age range of the M31 GC system resembled closely that of the Galactic GCs.  
However, a consistent and extensive set of UV photometry for the M31 GC system is an important 
next step in this effort. Here we report on the first imagery of the M31 GC system using the 
{\sl Galaxy Evolution Explorer} (GALEX) satellite (Martin et al. 2004).

In old stellar populations such as GCs, because of the relative longevity, hot He-burning 
horizontal-branch (HB) stars and their progeny are likely dominant UV sources (O'Connell 1999). 
Other sources of UV light in old populations are more difficult to model: extreme blue HB stars, 
post-asymptotic giant-branch (PAGB) stars, blue stragglers, and X-ray binaries.   
While metallicity is the dominant parameter controlling HB morphology in the oldest 
(Galactic halo-like) stellar populations, a number of studies suggest that age differences 
between clusters are the most significant second parameter affecting HB morphology 
(Lee, Demarque, \& Zinn 1994; Sarajedini, Chaboyer, \& Demarque 1997).
Population models also suggest that in a broad sense, FUV photometry may have the potential
to constrain age spreads within a  galaxy, and also between galaxies (e.g., Yi 2003).   
As our vision expands to consider more distant galaxies, the properties of the M31 and Galactic
GC systems play a significant role as templates because of the additional
detailed information that we have about the clusters in these systems.

\section{Observations and Data Analysis}

GALEX imaged 15 slightly overlapping fields of M31 covering a total $\sim 15$ deg$^{2}$
centered on M31; the mosaic covers most of the disk and halo of M31 (Bianchi et al. 2004; 
Thilker et al. 2004). The observations were obtained from 2003 September through October. 
The exposure times range from 650 to 1700 seconds, with each observation being obtained during 
the shadow crossing of one orbit. The satellite and on orbit performance of GALEX are described in 
Martin et al. (2004) and Morrissey et al. (2004).

Using the DAOPHOTII/ALLSTAR package (Stetson 1987), we performed aperture photometry and 
point spread function (PSF) fitting photometry for all detected objects in the GALEX M31 fields. 
Although the PSF-fitting was often successful, there is some variation of the PSF within 
a field in these early images, and stellar images are often elongated, because of the as-yet imperfect
removal of dithered pointing by image reduction pipeline. We consequently adopt 
aperture photometry for the final catalog. In order to deal with crowded fields, we use a 
4-pixel radius aperture and derive aperture corrections using isolated objects of moderate brightness. 
We found that the aperture corrections are stable at the 10\% level; our field-to-field variation across 
the M31 mosaic is also at the less than $10\%$ level. We expect this to improve in the final coadded data set.
Astrometry is obtained using objects for each field in common with the USNO-A2.0 catalog (Monet et al. 1998).  
The final photometry is on the AB magnitude system (Oke 1990).

We adopt the catalog of Barmby et al. (2000, hereafter B00) which gives positions for a 
large fraction of M31 GCs as well as optical magnitudes, reddening, and metallicity.  
Our final catalog (M. Rich and S.-C. Rey et al. 2004, in preparation) based on the fully co-added images 
will also match the full Bologna Catalog (Galleti et al. 2004).  For those clusters matching
to less than $5\arcsec$, we make subraster images (Figure 1) and verify the unique UV to optical match
by visual inspection. We reject those clusters falling more than 0.55 deg from the field center; 
the edge of the field has relatively poor positional accuracy and more severe image distortion.
Among the 435 cluster candidates of B00, we confirm 200 NUV and 94 FUV detections that have 
rms UV$-$optical positional agreement of less than $2\arcsec$.

Figure 2 shows the fraction of M31 GCs detected in the NUV and FUV bandpasses as a function
of $V$ magnitude and  $B-V$ color. Of the 384 GCs with both $V$ and $B-V$ data in B00,
190 (about 50\%) and 89 (about 23\%) GCs have been confirmed in NUV and FUV, respectively.
The color-magnitude diagram and color histogram show that most of the confirmed GCs
belong to blue clusters with $B-V < 1.2$.  In addition to lack of UV flux, a variety of
factors (e.g. artifacts and uncertain identification) may be responsible for the non-detection.
Our limiting magnitude for $NUV$ and $FUV$ is about 22 AB magnitude, which corresponds to 1 mag 
fainter than that of the UIT observations. We detected seven (NUV) and five (FUV) of the GCs with 
X-ray emission in Di Stefano et al. (2002). While the detections range from 
0.06 to 4.69$\times 10^{37} erg sec^{-1}$ we do not detect the two brightest X-ray clusters, 
Bo375 and Bo82.

Reddening corrections for individual clusters (including foreground and internal to M31) 
are based on a list of $E(B-V)$ kindly provided by P. Barmby.  We use the reddening law of
Cardelli, Clayton, \& Mathis (1989) for $R_{NUV}$ = 8.90 and $R_{FUV}$ = 8.16
(see also Bianchi et al. 1996). In our all analyses, we used only M31 GCs with $E(B-V)<0.15$,
which corresponds to the median value of the $E(B-V)$ distribution of M31 GCs (B00),
because uncertainty in the reddening and hence in the intrinsic colors
increases for large values of $E(B-V)$ (Barmby 2002).

\section{Results and Discussion}

How similar in broad-band colors are the M31 and Galactic GCs?
Figure 3 is the UV$-V$ colors versus $V$ diagram for M31 GCs. We compare both our sample and earlier 
UV satellite photometry, with dereddened optical magnitudes of B00. We convert magnitudes of 
the Galactic and UIT M31 GCs to those of the GALEX filter system using model spectral 
energy distributions for simple stellar populations (Yi 2003). We place the Galactic GCs 
on these plots using the M31 distance modulus of $(m-M)_0 =24.43$ (Freedman \& Madore 1990).   
The M31 and Galactic GCs occupy the same locus in these plots, and we find no systematic 
difference between UIT and GALEX photometry. Bohlin et al. (1993) reported a deficiency of 
analogs to the blue Galactic GCs with $(NUV-V)_0 \sim 3.2$ but our GALEX data
do not support their claim. Most GALEX M31 GCs populate $(NUV-V)_0 < 4.3$ and 
$(FUV-V)_0 < 5.5$, where we expect to find relatively UV bright clusters with bluer 
HB morphology.  We surmise that the red GCs with $(NUV-V)_0 > 4.3$ must have 
redder HB morphologies and are relatively UV faint, since these red clusters are probably 
metal-rich and UV$-V$ colors are very sensitive to metallicity.

Among their UIT sample, Bohlin et al. (1993) report blue clusters with $(NUV-V)_0 < 2.5$,
which are much bluer than most M31 and Galactic GCs. They suggest that these clusters are not
old GCs but instead are young clusters (see also Burstein et al. 1984; Bohlin et al. 1988; 
Cowley \& Burstein 1988). B00 also note that their GC catalog may be contaminated by 
other young objects with $B-V < 0.55$ and exclude 49 such objects from their analyses. 
Eight blue clusters of Bohlin et al. (1993) are included in our list of confirmed GCs 
({\sl filled circles}). Since $E(B-V)$ for these clusters is not available in B00, 
we assume that they are only affected by the foreground Galactic reddening of $E(B-V) = 0.10$ 
(Crampton et al. 1985). The arrows indicate reddening vectors by an increase of $E(B-V) = 0.10$. 
In our photometry, these clusters have bluer UV colors of $(NUV-V)_0 < 2.0$ and 
$(FUV-V)_0 < 3.0$, which are consistent with the color limits of blue clusters suggested by 
Bohlin et al. (1993). We will discuss the properties of these blue clusters in our forthcoming paper.

Figure 4 shows the distribution of our UV$-V$ colors versus [Fe/H], as an analog of the HB type versus 
[Fe/H] diagram (Lee et al. 1994; as such, bluer colors are now on the right), for M31 and Galactic GCs.  
The range in the UV$-V$ colors (compared to  $B-V$) is striking. While the M31 GCs 
extend over a wide range of $(NUV-V)_0$ color, the FUV sample is biased to the blue and metal-poor 
GCs with no redder [$(FUV-V)_0> 5.5$] clusters due to their faint $FUV$ magnitudes. 
We also superpose our three model isochrones for different ages ($\Delta t$ = -2, 0, and +2 Gyr),
which are constructed from our evolutionary population synthesis models of GCs in the GALEX filter 
system (see also Lee, Lee, \& Gibson 2002). The $\Delta t$ = 0 Gyr line ({\sl solid line}) corresponds to 
inner halo Galactic GCs (Galactocentric radius $\leq$ 8 kpc) at 12 Gyr. The $\Delta t$ = +2 Gyr 
({\sl long dashed line}) and -2 Gyr ({\sl short dashed line}) lines are for the models 2 Gyr older and younger 
than the Galactic GCs, respectively. These models include the treatment of the detailed systematic 
variation of HB morphology with age and metallicity and the contribution from PAGB stars. 
The models fit well the range in color of the clusters and their general locus on the 
age-metallicity plot. 

We find no metal-rich clusters with significant FUV flux, as is present in the two peculiar
Galactic GCs, NGC 6388 and NGC 6441 (Rich et al. 1997). These two clusters have hot HB stars 
that are seen in no other Galactic GCs of comparable metallicity (e.g., 47 Tuc) that have only red HBs.
The arrows represent the upper limit of the metal-rich clusters that were detected in NUV.   
There is a hint of dispersion at high metallicity in the $(NUV-V)_0$ versus [Fe/H] plot; 
if confirmed, this may reflect a milder version of the second parameter at high metallicity.

Many M31 GCs are resolved by HST, and for those we get HB morphology directly, which makes possible 
a comparison of the integrated properties of GCs with their actual resolved stellar populations. 
In order to see the UV flux variation with different metallicity and HB morphologies of GCs, 
we plot three different groups of GCs with known HB morphology from previous HST observations 
(Ajhar et al. 1996; Fusi Pecci et al. 1996; Rich et al. 1996; Holland, Fahlman, \& Richer 1997; 
Rich 2003): metal-poor blue HB (BHB) GCs with [Fe/H] $\sim$ -1.9 ({\sl filled circles}; G64 and G351), 
intermediate metallicity red HB (RHB) GCs with [Fe/H] $\sim$ -1.1 ({\sl filled squares}; G1 and G322), 
and metal-rich red clump GCs with [Fe/H] $\sim$ -0.6 (filled triangles; G58, G319, and G312). 
The sensitivity of metallicity and HB morphology to the UV flux is evident in both UV bands, 
i.e., the metal-poor BHB clusters show bluer UV$-V$ colors than the metal-rich red clump clusters, 
confirming the effect of the HB morphology on the UV$-V$ colors.

We also plot M2 and M72, obtained from the GALEX All-sky Imaging Survey observations. 
The exposure times were 129 and 112 sec for M2 and M72, respectively. The integrated $NUV$
and $FUV$ magnitudes were measured from the concentric large apetures (300$\arcsec$ and 120$\arcsec$ 
radius for M2 and M72, respectively) containing the total UV light of GCs. The error bar shows 
the magnitude variation due to the different aperture size (20 \% variation of aperture radius).
Although M2 ([Fe/H] $\sim$ -1.58; {\sl pentagon}) and M72 ([Fe/H] $\sim$ -1.54; {\sl hexagon}) 
have similar metallicities (Lee et al. 1994), they have different HB morphologies, in the sense that 
M2 has bluer HB morphology. They show significantly different $(FUV-V)_0$ colors 
[$\Delta(FUV-V)_0 \sim 1.11$] that are consistent with their different HB morphologies.

In the plot of $(NUV-V)_0$ color against [Fe/H], M31 and Galactic GC data appear to follow the trend 
indicated by the isochrones, a general correlation between $(NUV-V)_0$ color and metallicity.
On the other hand, comparison of the $(FUV-V)_0$ color with the models suggests the promise of gleaning 
age information (Yi 2003), provided that other sources of a small part of the scatter in Fig. 4 
(e.g., extreme HB, PAGB stars, and low mass X-ray binaries; King et al. 1993) can be quantified.
In the $(FUV-V)_0$ color versus [Fe/H] diagram, M31 and Galactic GCs are well located in the age range ($\pm2$ Gyr)
of model predictions and show no significantly different age distribution between two GC systems,
indicating that the M31 and Galactic GC systems show similar mean age and age spread, 
at least, for the oldest GCs.  This is consistent with claims for age spread in the Galactic and 
M31 GCs (Rosenberg et al. 1999; Barmby \& Huchra 2000; Salaris \& Weiss 2002). 
\\

The authors thank Pauline Barmby for kindly providing her catalog of M31 clusters. 
We are also grateful to Carla Cacciari for valuable comments.
GALEX (Galaxy Evolution Explorer) is a NASA Small Explorer, launched in 2003 April. 
We gratefully acknowledge NASA's support for construction, operation, and science analysis 
for the GALEX mission. Yonsei University participation is funded by the Korean Ministry of 
Science and Technology, for which we are grateful.

\clearpage

\begin{figure}
\centerline{\epsfysize=4in%
\epsffile{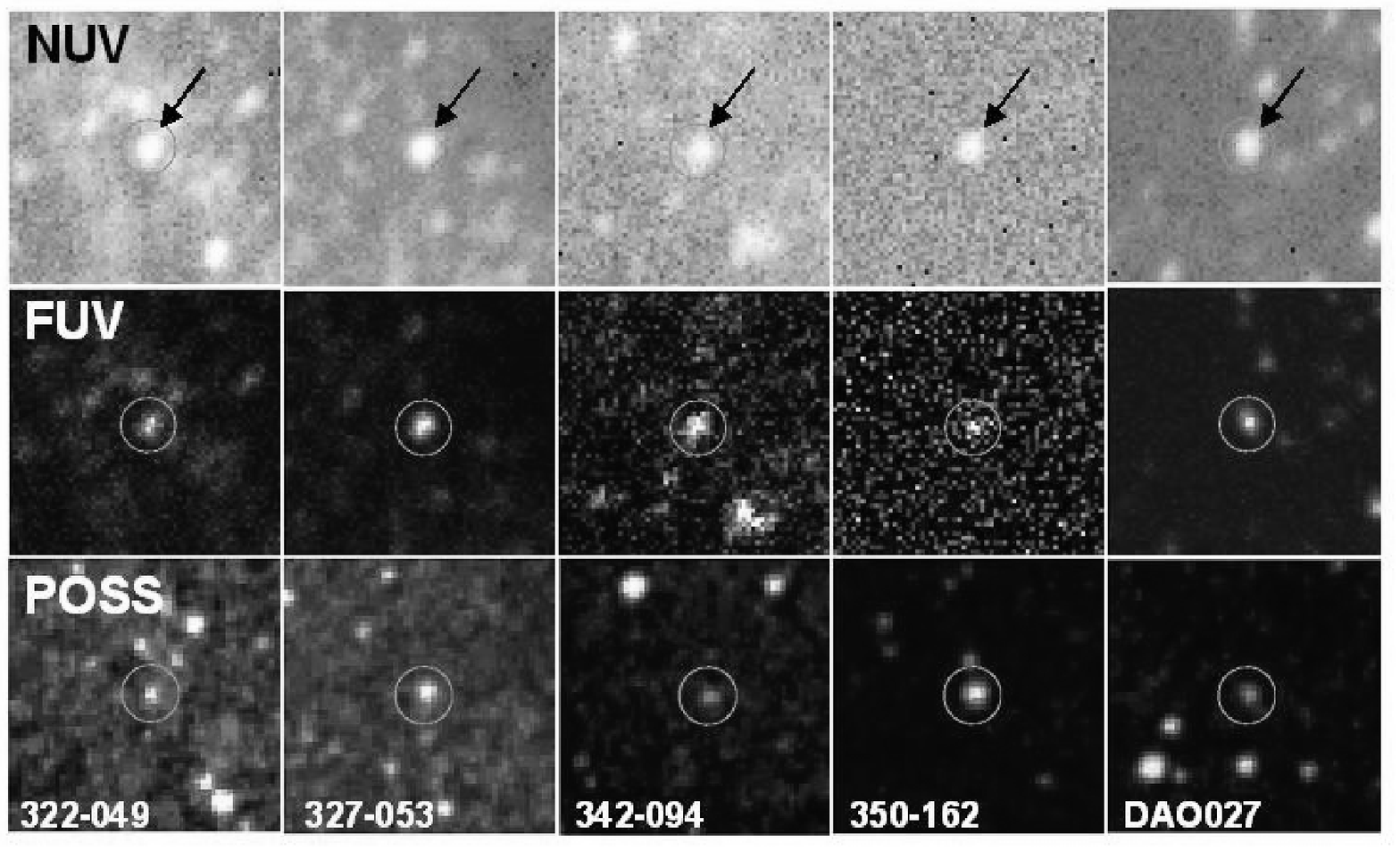}}
 \caption{GALEX images of confirmed M31 GCs for NUV ({\sl top}), FUV ({\sl middle}), and optical band images
from the Palomar Observatory Sky Survey ({\sl bottom}). All images are 90 arcsec$^{2}$.}
\end{figure}

\clearpage
\begin{figure}
\centerline{\epsfysize=7in%
\epsffile{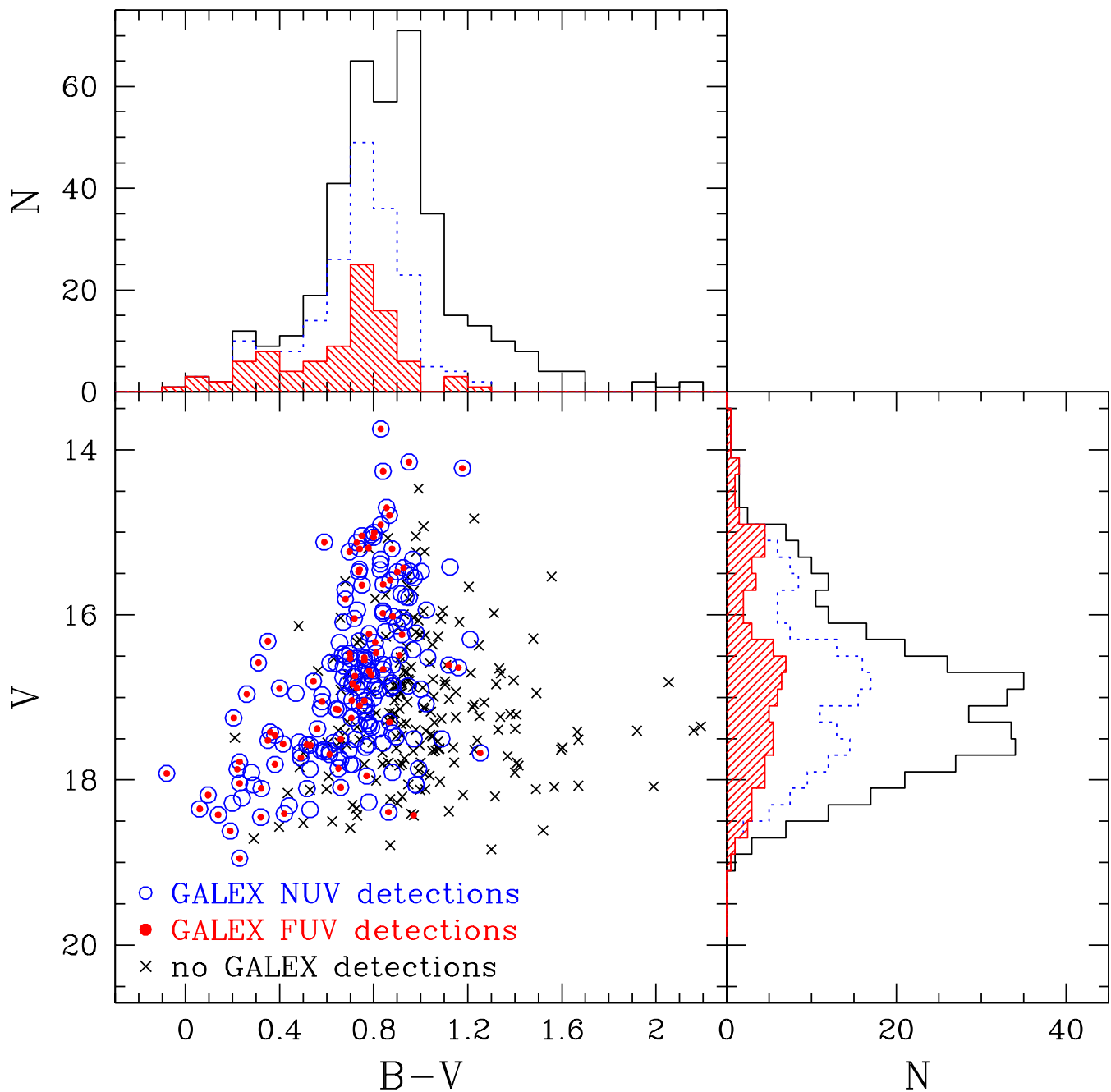}}
 \caption{Confirmed M31 GCs in the NUV and FUV bands.  Open and filled circles show GCs of Barmby et al. (2000)
confirmed in the GALEX NUV and FUV, respectively; crosses are not detected in the UV but are 
in the catalog of Barmby et al. The histograms are for the 384 clusters in Barmby et al. 
({\sl solid histogram}) and for the 190 and 89 clusters that are confirmed in NUV ({\sl dashed histogram}) 
and FUV({\sl shaded histogram}), respectively. As might be expected, we are more successful at detecting 
bluer clusters.}
\end{figure}

\clearpage
\begin{figure}
\centerline{\epsfysize=4in%
\epsffile{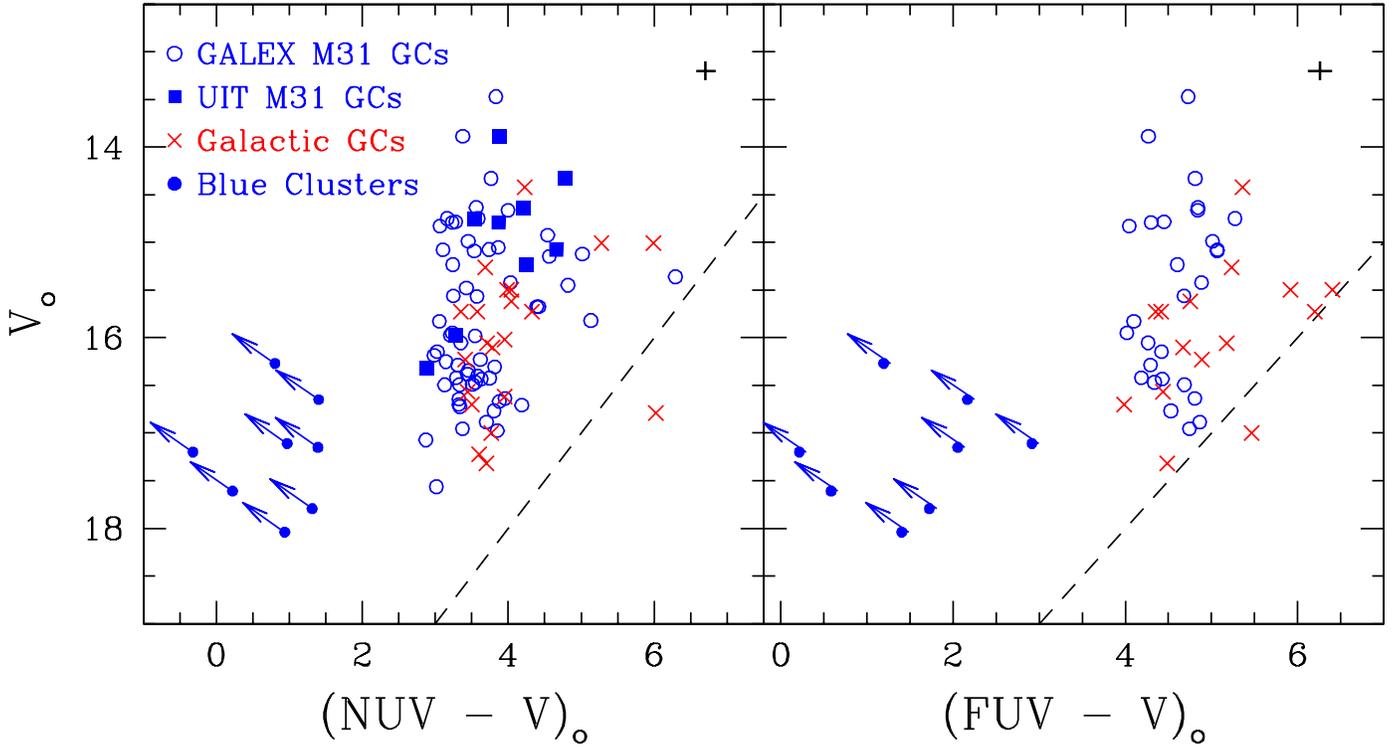}}
 \caption{UV$-V$ colors vs $V$ diagrams for GALEX M31 GCs with $E(B-V) < 0.15$. GALEX M31 GCs ({\sl open circles})
are compared with previous UIT results for M31 ({\sl filled squares}; Bohlin et al. 1993)
and Galactic GC results obtained from OAO-2 and ANS ({\sl crosses}; Dorman et al. 1995).
Young cluster candidates ({\sl filled circles}) are also shown with arrows indicating reddening vectors
by an increase of $E(B-V) = 0.10$ (see text). The sloping dashed line is our limiting magnitude. }
\end{figure}

\clearpage
\begin{figure}
\centerline{\epsfysize=4in%
\epsffile{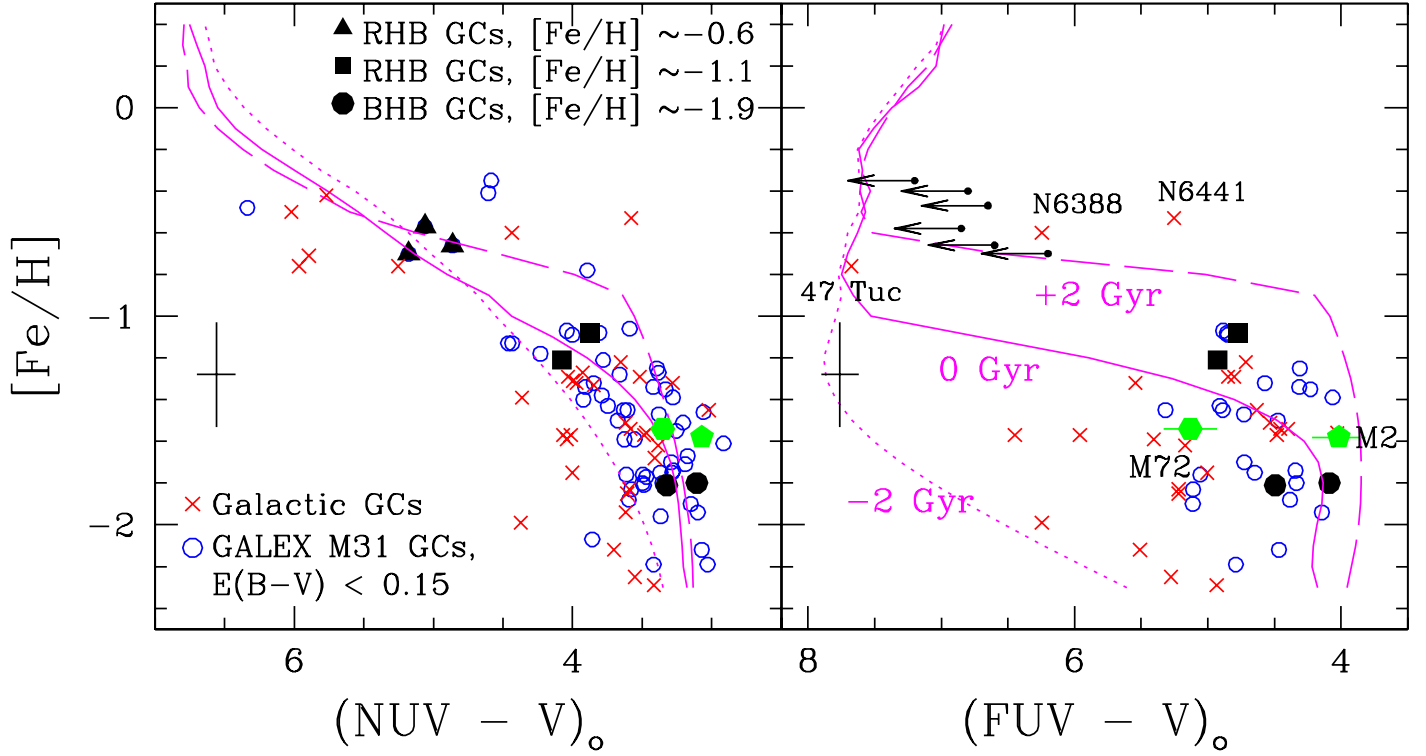}}
 \caption{UV$-V$ colors vs [Fe/H] diagrams for M31 ({\sl open circles}) and Galactic ({\sl crosses};
[Fe/H] from Harris 1996) GCs. The GALEX FUV sample is biased to the blue and metal-poor GCs 
because of our detection limit in FUV. Arrows indicate expected location of metal-rich 
([Fe/H] $>$ -0.8) M31 GCs in FUV; NGC 6388 and NGC 6441, which are metal-rich Galactic 
GCs with hot BHB stars (Rich et al. 1997), are indicated where they would be found if counterparts 
were present in M31. We plot three different groups of M31 GCs with known HB morphologies from 
previous HST observations. Two Galactic GCs, M2 ({\sl pentagon}) and M72 ({\sl hexagon}), obtained from GALEX
observations are also plotted. We also superpose our model isochrones indicating ages relative to 
the mean Galactic halo (see text). In $(FUV-V)_0$ vs [Fe/H], M31 and Galactic GCs are located in 
the age range ($\pm2$ Gyr) of model predictions and there is no indication for a systematic difference 
between M31 and the Milky Way. }
\end{figure}

\end{document}